\documentclass[doublecol,a4]{epl2}
\usepackage{graphicx}
\usepackage{dcolumn}
\usepackage{bm}

\title{Spin mixing in colliding spinor condensates: formation of 
an effective barrier
}

\author{M. Guilleumas\inst{1}, B. Juli\'a-D\'\i az\inst{1}, J. Mur-Petit\inst{2}, A. Polls\inst{1}}  
\shortauthor{M. Guilleumas, B. Juli\'a-D\'\i az, J. Mur-Petit, A. Polls}

\institute{
\inst{1} Departament d'Estructura i Constituents de la Mat\`{e}ria,
Universitat de Barcelona, E--08028 Barcelona, Spain\\
\inst{2} Department of Physics and Astronomy,
  University College London, 
  London WC1E 6BT,
  United Kingdom.
}


\abstract{
The dynamics of $F=1$ spinor condensates initially prepared 
in a double-well potential is studied in the 
mean field approach. It is shown that a small seed of $m=0$ 
atoms on a system with initially well separated $m=1$ and 
$m=-1$ condensates has a dramatic effect on their mixing dynamics,
acting as an effective barrier for a remarkably long time.
We show that this effect is due to the spinor character of the
system, and provides an observable example of the interplay between
the internal spin dynamics and the macroscopic evolution of the
magnetization in a spinor Bose-Einstein condensate.}

\begin{document}

\maketitle

\section {Introduction}

Ultracold atoms trapped by optical means are suitable systems 
to address a broad range of problems related to magnetic ordering 
and dynamics. Soon after the pioneering experiments with 
$F=1$ $^{23}$Na at MIT~\cite{Stenger1998} a number of groups  
have managed to observe spinor dynamics in a variety of conditions, 
e.g. in a quasi-1D system~\cite{Schmaljohann2004} or a 2D 
trap~\cite{ChangPRL2004}, thus being able to experimentally address 
many of the questions posed by 
theoreticians~\cite{HoLawOhmiKoashi,Lewens2007}.

Among the latter, the existing relation between the internal dynamics 
and the spatial ordering in Bose-Einstein Condensates (BEC) provides 
a beautiful example of how the microscopic interactions shape the 
macroscopic properties of the BEC. As discussed recently 
in~\cite{Saito2005, ZhangPRL05, Mur2006} one such example is the formation of 
magnetic domains in $F=1$ spinor systems (see also~\cite{Sadler2005}).
In Ref.~\cite{Mur2006} the dynamical evolution of a 1D confined BEC
with non-equilibrium initial populations in  
the spin components was studied in the mean field approach.
As the main emphasis of that study was
on domain  formation, the initial configurations considered were
always such that there was no relative motion between the center of
mass of the different Zeeman components.

In this paper we consider 
  the complementary case of
initial conditions where the spin components are spatially well
separated inside a harmonic trap, ensuring a certain amount of
relative center of mass motion between them.
The subject of collisions between scalar BECs has been addressed in
several experiments, revealing interesting features of collective
modes~\cite{LENS2000} and even realizing an atom analogue of 
the Hanbury Brown-Twiss effect of optics~\cite{IOTA2005}.
  Other experimental studies on the dynamics of spinor
  BECs~\cite{Miesner1999}, as well as binary
  mixtures~\cite{Hall1998,Lewandowski2002}, have focused on the 
  evolution of the spatial distribution of the different
  spin components in the system. Specially interesting is the
  observation by Hall {\em et al.}~\cite{Hall1998} of a fast
  convergence to a configuration with spin-segregation in the case of
  two interacting initially overlapping BECs of $^{87}$Rb  
  in different hyperfine states.

The setup considered here can be seen as a low energy confined version 
of those in Ref.~\cite{trip1}, where collisions of unconfined 
BECs are analyzed.
In the present work the system is confined and thus no atoms escape
the trap.

The paper is organized as follows. First the simplest extension 
of~\cite{Mur2006,Moreno2007} is considered, 
namely the formation of magnetic domains in a trapped $F=1$ BEC 
with no relative center of mass momenta between the Zeeman components
even when the center of mass momentum of the system is not zero.
Under these premises a full decoupling of the Kohn mode~\cite{Kohn1961}, 
dipolar oscillation in the confining trap, and the internal spin 
dynamics is found.

In the second step we introduce initial spatial separation between 
the different components, e.g. $m=-1$ on the right side of the trap and 
$m=+1$ on the left side, which are initially kept separated by means of 
a gaussian barrier. Switching off the barrier at $t=0$ the condensates 
collide and the time evolution of the different populations is analyzed. 

With these conditions we find a remarkable effect. Namely that 
on $F=1$ BECs with initially separated $m=-1$ and $m=+1$ components, 
a small amount of $m=0$ component produces a barrier-like effect 
which prevents the mixing of the $m=\pm 1$ components for a long
time. 
We further show that this effect is
characteristic of spinor BECs, in contrast e.g. to binary mixtures,
and occurs in a time scale reachable in current experiments 
with ultracold atoms.

\section {Description of the system}

In the mean-field approach, the $F=1$ spinor condensate is described
by a vector order parameter $\Psi$ whose components $\psi_m$ correspond 
to the wave function of each magnetic sublevel 
$|F=1, m \rangle \equiv |m\rangle$ with $m=1,0,-1$. In absence of an
external magnetic field and at zero  
temperature the spin dynamics of this system confined in an external 
potential, $V_{\textrm{ext}}$, is described by the following coupled
equations for the spin components~\cite{HoLawOhmiKoashi,Mur2006}:
\begin{eqnarray}
i \hbar \, \partial \psi_{\pm 1}/\partial t &=&
 [{\cal H}_s  + c_2(n_{\pm 1}+n_0- n_{\mp 1})] \, \psi_{\pm 1}  \nonumber \\
 &&+ c_2 \, \psi_0^2 \psi^{*}_{\mp 1} \,, \label{dyneqs0} \\
i \hbar \, \partial \psi_0/\partial t &=&
 [{\cal H}_s  + c_2(n_{1}+n_{-1})] \, \psi_{0} \nonumber \\
   &&+ c_2 \, 2 \psi_{1} \psi_0^* \psi_{-1} \,, \label{dyneqs1}
\end{eqnarray}
with 
${\cal H}_s=-\hbar^2/(2M)\, {\bm \nabla}^2 +V_{\textrm{ext}}+c_0n$
being  the spin-independent part of the Hamiltonian. 
The density of the $m$-th component is given by 
$n_{m}({\bf r})=|\psi_m({\bf r})|^2$, 
while $n({\bf r})=\sum_m|\psi_m({\bf r})|^2$ is the total density
normalized to the total number of atoms $N$. The population of each
hyperfine state is  
$N_m=\int d{\bf r} |\psi_m ({\bf r})|^2$. Defining the relative populations 
$\lambda_m=N_m/N$, it follows that $\lambda_1+\lambda_0+\lambda_{-1}=1$ and 
the magnetization of the system is 
${\cal  M}=\lambda_1-\lambda_{-1}$. The total number of 
atoms and the magnetization are both conserved quantities~\cite{Mur2006}.
The couplings are $c_0=4\pi\hbar^2(a_0+2a_2)/(3M)$ and
$c_2=4\pi\hbar^2(a_2-a_0)/(3M)$, where $M$ is the atomic mass and 
$a_0,a_2$ are the scattering lengths describing binary elastic
collisions in the channels of total spin 0 and 2, respectively.
The interatomic interactions permit the transfer of population
between the different Zeeman components by processes that conserve
the total spin, 
$|0\rangle + |0\rangle \leftrightarrow |+1\rangle + |-1\rangle$.

\section{Preparation of the system}

We consider $N = 20000$ atoms of spin-1 $^{87}$Rb in a highly
elongated trap with $\omega_{\perp} = 2 \pi \times 891$~Hz  and 
$\omega_z = 2 \pi \times 21$~Hz~\cite{Schmaljohann2004}.
The scattering lengths are $a_0=101.9 a_B$ and $a_2=100.4 a_B$ 
corresponding 
to a ferromagnetic behavior, $c_2<0$~\cite{HoLawOhmiKoashi}.
Since $\omega_{\perp} \gg \omega_z$ the dynamics takes place along the
axial direction and the equations of motion become one-dimensional
for the longitudinal wave functions  $\psi_m(z)$ by rescaling the
coupling constants $c_0$ and $c_2$ by a factor $1/(2\pi a_{\perp}^2)$,
with $a_{\perp}$ the transverse oscillator
length~\cite{Moreno2007,Zhang2005a}.
Overlapped with the harmonic potential we consider a gaussian barrier
which separates the system in two symmetric wells (labeled as $R$ and $L$,
corresponding to the right and left side respectively).
The resulting confining potential is the same for the three 
Zeeman components, and  reads:
\begin{equation}
V_{\textrm{ext}}=\frac{m}{2} \omega_z^2 z^2 + A \;{\exp}(-z^2/\sigma^2) \,.
\end{equation}

Given an initial configuration of the system (the relative 
population of spin components inside each potential well:
$\lambda_m^j$, with $m=1,0,-1$, $j=L, R$ and 
$\lambda_m =\lambda_m^L+\lambda_m^R$), 
the initial state of the system is prepared as follows: 
(i) First the ground-state wave function of the scalar condensate is
calculated from ${\cal H}_s$ on the whole spatial domain. The presence 
of a wide and large enough barrier at the center of the trap ensures
no overlap between the wave function at each side of the
barrier. Notice that due to the presence of the initial
barrier the two wells are not parabolic and the wave function in each
well is strongly non-parabolic, see inset of Fig~\ref{fig2}.
(ii) Afterwards, the wave function of each magnetic sublevel inside each 
well is obtained by normalizing the scalar wave function to the desired 
initial relative population $\lambda_m^j$ in that well~\cite{Moreno2007}.
(iii) At $t=0$ the potential barrier is instantaneously switched off
and the spinor system is allowed to evolve inside the harmonic 1D-potential.
The dynamics follows by solving the system of coupled equations 
(\ref{dyneqs0},\ref{dyneqs1}). Our numerical procedure for the 
time evolution combines the split operator method with the fast 
Fourier transform to treat the kinetic terms and a fourth-order 
Runge-Kutta method for the remaining terms of the 
dynamical equations~\cite{Mur2006}.

\section{Scalar condensate}

In order to understand the spinor-driven effects on the condensate
dynamics, let us first address 
a spin-polarized condensate of $m=-1$ atoms all of
them localized  on the right side of the trap: $\lambda_m^L$=0 (for
all $m$), $\lambda_1^R=\lambda_0^R=0$ and $\lambda_{-1}^R=1$.
Since all the atoms are in the same hyperfine state this case 
is equivalent to having a scalar condensate. At $t=0$ the gaussian 
barrier is switched off and the system evolves inside the harmonic 
trap bouncing from right to left in the well-known Kohn
mode~\cite{Kohn1961}.

\begin{figure}[t]
\includegraphics[width=0.9\columnwidth,angle=0, clip=true]{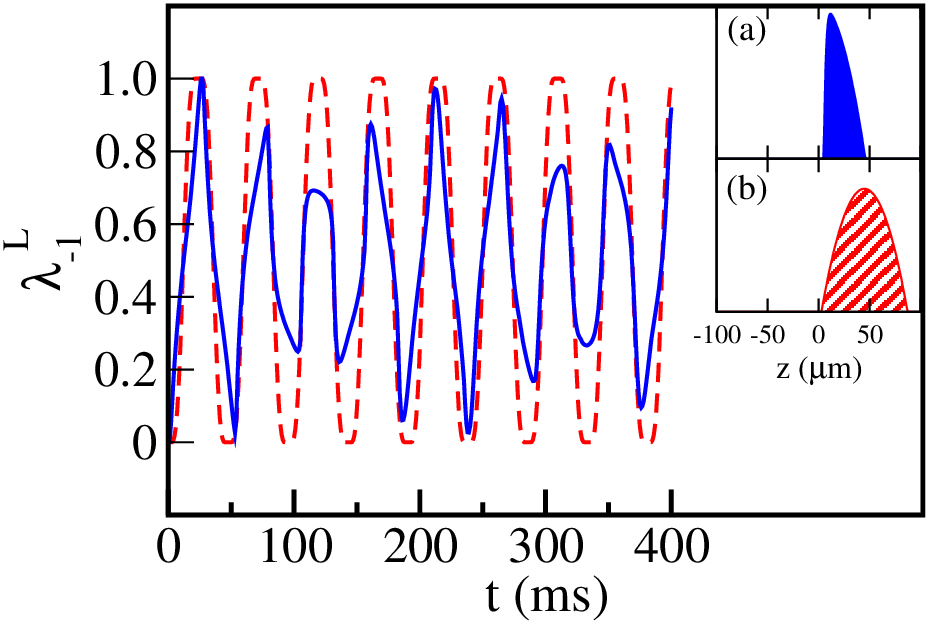}
\caption[]{Fraction of atoms on the left side 
  of the trap as a function of time for two different initial
  states of a scalar BEC (sketched in the inset): (a) harmonic trap with a
  gaussian barrier (solid line) and (b) ground state of a displaced
  harmonic trap (dashed).}
\label{fig1}
\end{figure}

In Fig.~\ref{fig1} we depict the fraction of atoms in
the left side of the trap as a function of time (solid line). For comparison, the
dipole oscillation of the ground state of a displaced harmonic 
trap is also displayed (dashed line). The Kohn mode is clearly
identified by the frequency of the population oscillations being equal
to the axial frequency of the trap $\omega_z$. 
Further, as our initial configuration is not the ground state of the
harmonic trap, the Kohn oscillation is convoluted with 
other frequencies associated with the deformation of the density
profile that bounces inside the trap.

\section{Component driven dynamics}

We proceed now to introduce spin dynamics in the system. 
To this end, we analyze the time evolution of a system with all 
three Zeeman components populated, and 
the atoms initially in the right side of the trap.

If we start with $\lambda_{\pm1}^R=25\%$ and $\lambda_0^R=50\%$
(and $\lambda_m^L=0~\forall m$), which corresponds to the ground-state
configuration in spin space~\cite{Mur2006}, we observe no transfer of 
population among the different components, while the total density follows
the same behaviour as the solid line in Fig.~\ref{fig1} (Kohn mode).
Moreover, if we start from a spin configuration away from the 
ground state, the total density still follows a dipole-like
oscillation, on top  
of which a transfer of population happens identical to that occurring 
in the case of a system at rest in a simple harmonic trap~\cite{Mur2006,peli}.
In other words, the center-of-mass motion and the internal (spin) 
dynamics are fully decoupled, as expected for the dipole mode in a harmonic 
trap~\cite{kohn-harmonic}. We remark that the time evolution of $\lambda_m(t)$ 
is exactly the same as in the system
at rest. The relative populations oscillate around the corresponding
equilibrium values at a given magnetization 
(e.g., $\lambda_0=50\%$, $\lambda_{\pm 1}=25\%$ for ${\cal M}=0$), 
as was already pointed out in Ref.~\cite{Moreno2007}.

We have seen how the presence of internal spin dynamics does not affect 
the evolution of the total density $n(z,t)$ of the system when all the atoms 
have the same initial spatial distribution. Now we will illustrate how 
the exact initial conditions of the system (spin configuration and 
spatial distribution thereof) can have a strong effect on the internal and the spatial dynamics,
showing genuine effects associated to the spinor nature of the
condensate.

\begin{figure}[tb]
\includegraphics[width=0.9\columnwidth,angle=0, clip=true]{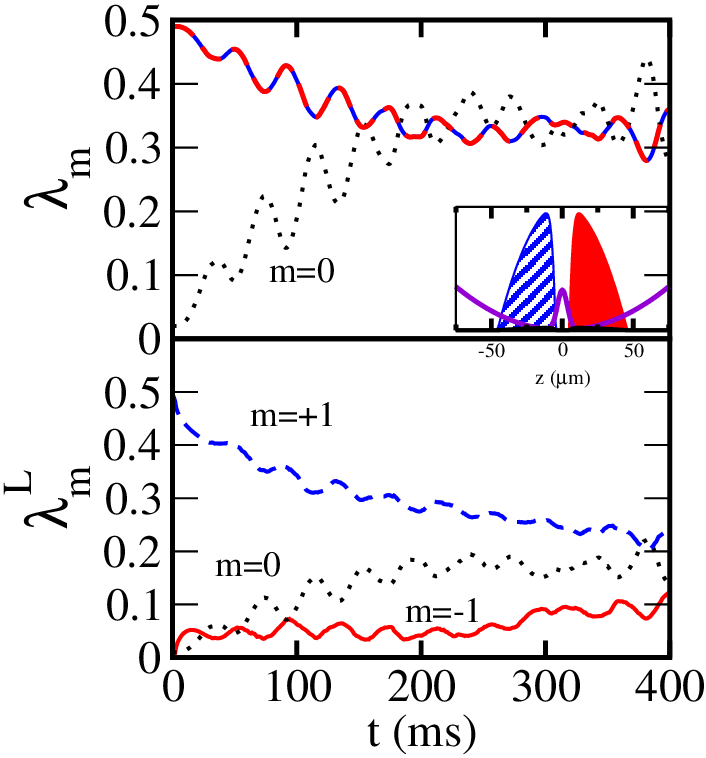} 
\caption[]{Time dependence of the total (top) and fraction on 
  the left side (bottom) relative populations of the various Zeeman 
  components of a spinor condensate, for the initial conditions
  $\lambda_1^L=\lambda_{-1}^R=49\%,~\lambda_0^R=\lambda_0^L=1\%$.
  Solid, dotted  and dashed correspond to $m=-1$, $m=0$ and $m=+1$
  respectively. Notice that the lines for $m=\pm1$ in the top panel
  are indistinguishable.} 
\label{fig2}
\end{figure}

Let us consider the case where initially all $m=1$ $(m=-1)$ atoms are
located on the left (right) side of the trap, with a vanishing overlap
between the two condensates
($\int_{-\delta}^{+\delta} {\rm d} z [n_{-1}(z) +  n_{+1}(z)]=0$ for  
$\delta=4\mu$m).
A small amount of atoms with $m=0$ is added on both 
sides, $\lambda_0^L=\lambda_0^R=1 \%$, together with
$\lambda_1^L=\lambda_{-1}^R=49 \%$ and
$\lambda_{-1}^L=\lambda_1^R=0$ (see inset of Fig.~\ref{fig2}).
This small admixture of $m=0$ allows
the exchange of atoms between spin components:
$|0\rangle + |0\rangle \leftrightarrow 
|+1\rangle + |-1\rangle$ within the approach of
Eqs.~(\ref{dyneqs0},\ref{dyneqs1}). Then, the population of each sublevel 
changes with time but due to the symmetry of the equations
and the initial conditions we have that for any $t$, 
$\lambda_1(t)=\lambda_{-1}(t)$ (see Fig.~\ref{fig2}). 

The resulting dynamics after switching off the trap is as follows.
At short times, $m=0$ atoms are created at the center of the trap
where the overlap between $m=\pm 1$ components is larger
[cf.~Eq.~(\ref{dyneqs1})]. Here the $m=\pm 1$ components fuse together
giving rise to $m=0$ atoms with a conversion rate proportional to
$c_2$. This can be seen in the lower panel of Fig.~\ref{fig2}, where
we present the relative populations on the left side of the trap as a 
function of time. We observe after $\sim 5$~ms  
a temporary saturation of the number of $m=-1$ atoms on the left side
for up to 350~ms, while the $m=+1$ component keeps losing population,
which is transferred to the $m=0$ Zeeman state.
The oscillations seen in all $\lambda_m^j$ in Fig.~\ref{fig2} are
reminiscent of the Kohn mode discussed above, and have a dominant
frequency close to that of the harmonic trap ($\omega_z$). 

The long time behaviour observed in Fig.~\ref{fig2} is determined 
by  the initial conditions that correspond to a highly excited configuration.
After the clouds move towards each other and collide, the initial collision energy is 
distributed among internal degrees of freedom, thus acting as an effective 
temperature. Therefore the populations perform 
damped oscillations around the equipartition configuration \cite{Moreno2007}.

\begin{figure}[t]
\includegraphics[width=0.7\columnwidth,
angle=0, clip=true]{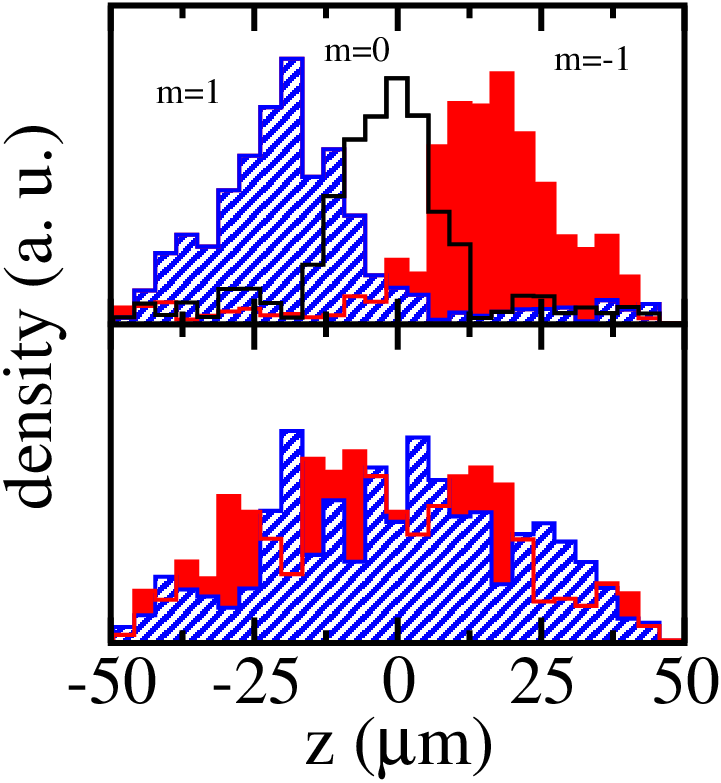}
\caption[]{Density distribution of atoms in the trap, at time $t=115$~ms  with
$\lambda_0(t=0)=2\%$ (top) and for the case of 
  $\lambda_{0}(t=0)=0$ (bottom). Solid, empty and dashed correspond to $m=-1$, 
$m=0$ and $m=+1$. A $\sim 4\, \mu$m spatial binning has been employed 
to generate the plot.
\label{fig3}}
\end{figure}

To further understand the effect of the $m=0$ component, we show in
the upper panel of Fig.~\ref{fig3} the density profile of the various Zeeman
components at $t=115$~ms, well into the time evolution of the system.
We observe the presence of $m=0$ atoms located at the 
center of the trap, with
$m=1$ ($m=-1$) atoms remaining to its left (right).
In summary, the initial $m=\pm1$ components are steadily converted
into $m=0$ atoms which stay mostly at the center of the trap, and act
as an effective barrier that prevents the spatial mixing of the
surviving $m=\pm1$ components. These stay mostly on their original side of the
trap for several hundreds of miliseconds, with only a small fraction
of atoms crossing over to the other side.  
The long lifetime that the $m=0$ barrier shows in this case is to
be contrasted with the much faster spin dynamics seen previously,
where the typical timescale for population transfer between 
 initially overlapping Zeeman states was $\sim$ 100~ms~\cite{Mur2006}. 
It also contrasts with the spin-segregation time scale observed 
in binary mixtures of $^{87}$Rb~\cite{Hall1998}, as will be 
addressed below. Interestingly, qualitatively similar results 
have been obtained in an antiferromagnetic system simulated by changing the 
sign of $c_2$ ($c_2>0$). 

We have also observed that even a very small $m=0$ seed in one side $(0.02 \%)$ has
the same effect of acting as a barrier and stops the mixing of the
$m=\pm 1$ atoms. However, as expected, different $\lambda_0(t=0)$
result in  different transient behaviours. In particular, the larger
the $m=0$ seed the longer the time necessary to have a mixture of the
other Zeeman components with similar concentrations in both sides.
Nevertheless, the long time ($\sim 400$~ms) configuration is
roughly the same.

To emphasize the spinorial origin of this barrier effect we compare
with the mixing process of two interacting BECs. 
Such binary mixtures have been realized experimentally by coupling two Zeeman
states via a two-photon drive. For example, mixtures of $^{87}$Rb in 
the hyperfine states $|F=2,m=2\rangle$ and $|2,1\rangle$~\cite{LENS2000} as 
well as $|1, -1\rangle$ and $|2,1\rangle$~\cite{Hall1998,Lewandowski2002} 
have been both produced.
In the mean-field framework their dynamics can be easily simulated 
with the coupled differential equations~(\ref{dyneqs0},\ref{dyneqs1})
by taking a zero amount of $m=0$ atoms.

\begin{figure}[t]
\includegraphics[width=0.9\columnwidth,
angle=0, clip=true]{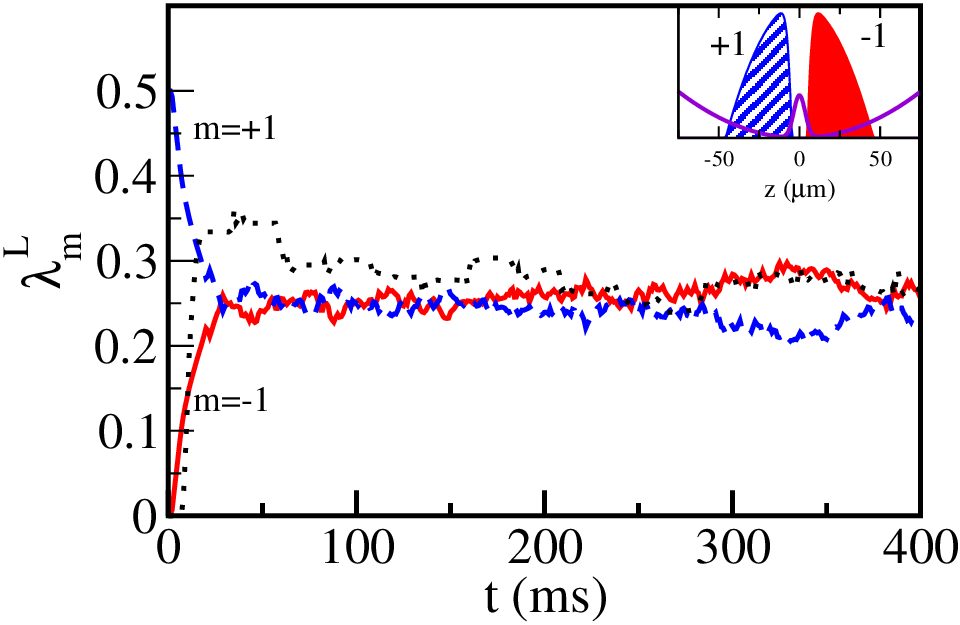}
\caption[]{Fraction of atoms in the $m=-1$ (solid) 
and $m=+1$ (dashed) Zeeman component on the left side of the trap, 
as a function of time for a binary mixture. The inset shows the 
initial spatial distribution
together with the confining potential before switching off the
$\sigma=5 \mu$m gaussian barrier.
The dotted line stands for the time evolution of $\lambda_{-1}(t)$ for 
an initial configuration given by $\sigma=40\mu$m.}
\label{fig4}
\end{figure}

The preparation of the system follows the same steps as above,
setting now the initial conditions as
$\lambda_{+1}^L=\lambda_{-1}^R = 50\%$
and $\lambda_{+1}^R=\lambda_{-1}^L = 0$.
The barrier is switched off at $t=0$ and the two condensates move
inside the harmonic trap towards each other, colliding after 
$\sim 10$~ms. As can be seen in Fig.~\ref{fig4}, at this time a sizeable 
amount of atoms of each component has already reached the other side 
of the trap, and rapidly mixes with the other component. 
As expected, $\lambda_{-1}^L$  grows initially with
a similar rate as it did in Fig.~\ref{fig2}, but departs from that
behaviour after times $\sim$ 5~ms, due to atoms coming from the other
side of the trap. There is also a similar decrease of $\lambda_{+1}^L$
since an equal amount of $m=1$ atoms goes to the right.
Around $t=25$~ms the two components are mixed and there is almost the
same number of $m=\pm 1$ atoms in each side of the trap, i.e. there is
a fast and complete mixing of both components, as shown in the lower
panel of Fig.~\ref{fig3}.

If the width of the initial gaussian barrier is increased, the initial
collision energy is larger, and thus the number of \nobreak{$m=-1$}
atoms reaching the left side is also larger initially. However, it
reaches the same asymptotic value as before (cf.~dotted line in
Fig.~\ref{fig4}). Therefore, we conclude that the transition to a 
steady state is much slower in the spinor case than it is for a 
binary mixture~(cf.~Figs.~\ref{fig2} and~\ref{fig4}), which 
qualitatively agrees with the observations in~\cite{Hall1998,Miesner1999}.

\section{Conclusions}

In summary, we have studied the interplay between the internal spin
dynamics and the spatial evolution of a spin-1 condensate  in a
confining potential. We have described the effect produced by the
presence of $m=0$ atoms in a system initially prepared with two
spatially separated spin components $m=+1$ and $m=-1$. 
An initial small population of the $m=0$ component drastically 
affects the mixing time of the system as compared to a binary mixture,
producing a barrier-like behaviour which keeps the other two
components from mixing for times up to $\sim 350$ ms which is a 
timescale on which experimental observations should be feasible. We 
have also demonstrated the genuine spinor character 
of this effect, and that it does not appear in scalar condensates 
or binary mixtures. The population transfer term 
of Eqs~(\ref{dyneqs0},\ref{dyneqs1}) plays a key role in the 
generation of the effective barrier.

The seed of $m=0$ atoms can be a residual of the preparation of the
initial state in an experiment. A similar effect would also be
produced in an initially pure,
$|+1\rangle\,|-1\rangle$ mixture by two-body spin-flip
processes $|+1\rangle + |-1\rangle \rightarrow |0\rangle + |0\rangle$
at the initial stages of the evolution.
Therefore, the presence of the effective barrier preventing the mixing 
of spin components appears as a general phenomenon in spinor BECs 
prepared with spatially separated Zeeman components.

In the future, we will address more general examples of interplay of
spin and spatial degrees of freedom in spinor Josephson junctions
whose barriers have a more complex time dependence, and can
distinguish between different hyperfine components~\cite{Thorn2008}.
We note also that more exotic spatial orderings and dynamics are
expected to occur in optical lattices, where oscillation in the
populations of two different Zeeman components have already been
reported~\cite{TrotzkyScience2008}.

\acknowledgments
It is a pleasure to acknowledge K. Bongs, M. Lewenstein and
L. Pitaevskii for a careful reading of an earlier version 
of the manuscript and also 
K. Sengstock for useful discussions. B.J-D. is supported 
by a 'Juan de la Cierva' contract, MEC/Spain. J.M.-P. acknowledges 
support from EPSRC (UK) and QUDEDIS (ESF). This work is also
partially supported by Grants No. FIS2005-03142 and FIS2005-01414 from
MEC (Spain) and 2005SGR-00343 from Generalitat de Catalunya.

\end{document}